\documentclass[twocolumn,floats,prl,superscriptaddress,showpacs,aps]{revtex4}
\usepackage[dvips]{graphicx}
\usepackage{textcomp}

\begin{document}

\title{Local tunneling spectroscopy of the electron-doped cuprate Sm$_{1.85}$Ce$_{0.15}$CuO$_4$}

\author{A. Zimmers}
\email{Alexandre.Zimmers@insp.jussieu.fr}
\affiliation{Institut des
Nanosciences de Paris, CNRS UMR 7588, Campus Boucicaut, 140 rue de
Lourmel, F-75015 Paris, France} \affiliation{Center for
Superconductivity Research, Department of Physics, University of
Maryland, College Park, Maryland 20742, USA.}

\author{Y. Noat}
\author{T. Cren}
\author{W. Sacks}
\author{D. Roditchev}
\affiliation{Institut des Nanosciences de Paris, CNRS UMR 7588,
Campus Boucicaut, 140 rue de Lourmel, F-75015 Paris, France}

\author{B. Liang}
\author{R. L. Greene}
\affiliation{Center for Superconductivity Research, Department of
Physics, University of Maryland, College Park, Maryland 20742, USA.}

\begin{abstract}
We present local tunneling spectroscopy in the optimally
electron-doped cuprate Sm$_{2-x}$Ce$_{x}$CuO$_4$ x=0.15. A clear
signature of the superconducting gap is observed with an amplitude
ranging from place to place and from sample to sample
($\Delta\sim$~3.5-6meV). Another spectroscopic feature is
simultaneously observed at high energy above $\pm$50meV. Its
energy scale and temperature evolution is found to be compatible
with previous photoemission and optical experiments. If
interpreted as the signature of antiferromagnetic order in the
samples, these results could suggest the coexistence on the local
scale of antiferromagnetism and superconductivity on the
electron-doped side of cuprate superconductors.
\end{abstract}


\maketitle Due to strong correlations, doping a Mott insulator
yields a number of fascinating properties in a great number of
materials. Among these, understanding the properties of cuprates
remains one of the greatest challenges in condensed matter physics.
In recent years, an important effort has been made to understand the
interplay between the antiferromagnetic (AF) and the superconducting
orders. A close relation between these two is undeniably present on
the electron doped side of the phase diagram since both orders
overlap in a a narrow doping range. This observation was first
reported by angle resolved photoemission (ARPES) when revealing the
presence of a large energy gap on parts of the Fermi surface
\cite{ArmitageNCCO} in superconducting samples. The particular
symmetry of this large energy gap, which happens at the interception
between the nominal Fermi surface and the antiferromagnetic
Brillouin zone, has led to its interpretation as the signature of AF
order in the sample. The characteristic temperature T' below which
this large energy gap opens has since been mapped out over the phase
diagram by infrared spectroscopy \cite{Onose, Zimmers}, revealing
its presence up to x=0.17 in superconducting PCCO samples. However,
neither ARPES nor infrared spectroscopy are able, in principle, to
differentiate if this large energy gap is due to long range AF order
or AF fluctuations. These technics can only estimate at best the
minimum fluctuation length and time scales involved. A recent
inelastic neutron diffraction measurement \cite{Greven} has resolved
this issue by showing that in superconducting samples, the
antiferromagnetic correlation length does not diverge but saturates
at low temperatures. This suggests that only AF fluctuations are
present with a characteristic length scale of 200\AA\, in optimally
doped samples. An important challenge is now to see whether both
orders do coexist on a local scale.

To answer this question, we report a first set of local tunneling
spectroscopy measurements in STM geometry on the electron-doped
cuprate Sm$_{1.85}$Ce$_{0.15}$CuO$_4$ (SCCO) on larger energy
scales. The tunneling conductance spectra show simultaneously the
superconducting gap at low energies and a high energy feature at
same spatial positions. The attribution of this high energy feature
to the presence of local AF order suggests that, in electron-doped
cuprates, antiferromagnetism and superconductivity coexist locally
at optimal doping. In addition to these various energy scales, our
measurements have revealed no changes in the superconducting gap
magnitude within a 330\AA\, map but show large variations of this
gap (up to 70\%) between distant points within a same sample and
between two different optimally doped samples. These large
variations could be explained by modulation of the cerium and oxygen
content within the samples. This observation is different from
previous scanning tunneling microscopy (STM) measurements in the
hole-doped cuprate Bi-2212 which show local inhomogeneities in the
superconducting gap on the nanoscale \cite{Cren, Davis}.

Local electronic properties are probed by scanning tunneling
microscopy/spectroscopy (STM/STS), where the tunneling current $I(V
)$ is measured between an atomically sharp tip and a sample as a
function of the bias voltage. Images of the surface can be made by
scanning the tip (topographic mode) or at a given point
(spectroscopic mode). The tunneling junctions were achieved by
approaching mechanically cut Pt/Ir tips (whose density of states
(DOS) near the Fermi level is roughly constant) to the c-axis
oriented surface of the crystals. This configuration is believed to
be most sensitive to the DOS of the $CuO_2$ planes of SCCO. The
differential conductance measured, $dI/dV$, is proportional to the
sample local DOS convoluted with the Fermi function (at the energy
$E_F$ + eV).

\begin{figure}
\begin{center}
\includegraphics[width=9cm]{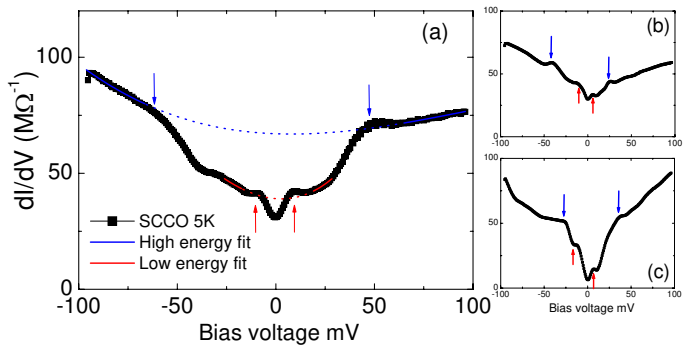}
\end{center}\caption{(Color online) (a) Raw spectra  at 5K of SCCO x=0.145 (onset $T_c$=20K).
A low and high energy features are identified by the red and blue
arrows. The red and blue fits were used in figure \ref{Fig2} (a)
and (b) to divide out the background. For the high energy
background, the fitting region was chosen to be [-180;-120]meV and
symmetrically from [120,180]meV. For the low energy background the
fitting region was chosen from [-55;-35]meV and from [35;55]meV.
The intermediate region in both fits (blue and red dashed lines)
were not used when evaluating the background fit but were used
when normalizing the spectra of figure \ref{Fig1} to obtain
figures \ref{Fig2} (a) and (b).\\ (b) and (c) Raw spectra on the
same sample but at distant positions. The superconducting gap is
always centered around zero bias as expected, however we observe
that the high energy feature can be shifted to negative (panel
(b)) or to positive bias (panel (c)) depending on the
location.}\label{Fig1}
\end{figure}

\begin{figure}[t]
\begin{center}
\includegraphics[width=6.7cm]{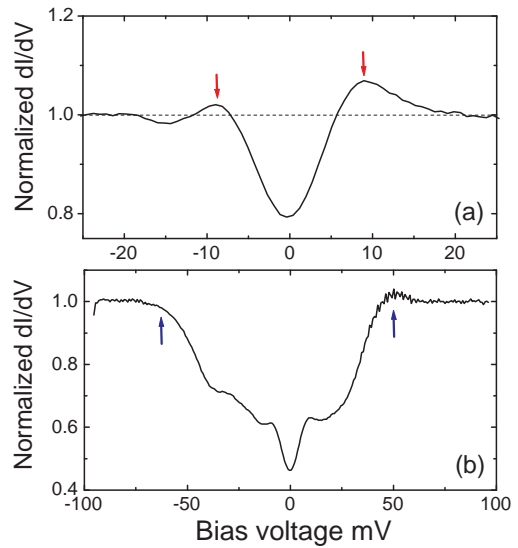}
\end{center}\caption{(Color online) (a) Low energy feature isolated by dividing out the low energy
background (red fit in figure \ref{Fig1}). This feature is
identified as the superconducting gap and is found to have a
magnitude of $2\Delta$=12meV in a BCS picture.\\(b) High energy
feature isolated by dividing out the high energy background (blue
fit figure \ref{Fig1}). The high energy feature magnitude is
identifiable by the blue arrows.} \label{Fig2}
\end{figure}

Single crystals of Sm$_{2-x}$Ce$_{x}$CuO$_4$ were grown by flux
method. Samples were typically a few mm along a and b axis and a few
hundred microns along c axis. Samples were annealed at
925\textcelsius\, under reduced oxygen environment for 3 to 5 days.
Ce concentrations were determined by WDX. Superconducting transition
temperature were determined by DC susceptibility. The two samples
presented here are SCCO with cerium concentrations x=0.145$\pm$0.005
with an onset $T_c$=20K and SCCO x=0.155$\pm$0.005 with an onset
$T_c$=17.5K. These concentrations are just an indications of the
carrier density in the samples. Indeed the sample carrier density
can also slightly change due to various oxygen reduction treatments.
These different treatments could, for instance, explain the small
differences in the measured critical temperatures \cite{Higgins}
even though both samples are near optimally doped in cerium. These
materials were chosen since they present the advantage of being
cleavable compared to other electron-doped cuprates such as
Nd$_{2-x}$Ce$_{x}$CuO$_4$ (NCCO) and Pr$_{2-x}$Ce$_{x}$CuO$_4$
(PCCO).

\begin{figure}
\begin{center}
\includegraphics[width=6.5cm]{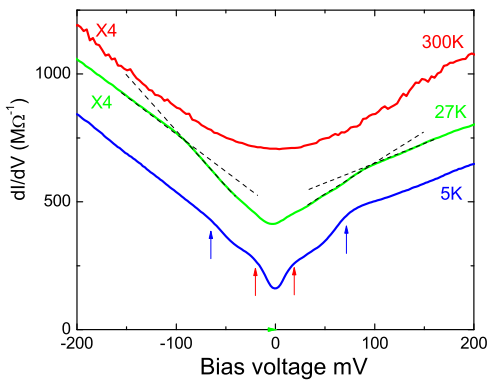}
\end{center}\caption{(Color online) Temperature evolution of the
tunneling spectra of SCCO x=0.155 (onset $T_c$=17.5K) at 5K, 27K and
300K (the 27K and 300K curves are multiplied by 4 for clarity). The
superconducting gap seen at 5K (see red arrows) is found to be
closed at 27K which is above $T_c$. The high energy feature is
however still present (see dash lines as a guide for the eye). The
high energy feature is found to be closed at room temperature. Using
previous optical results the high energy characteristic temperature
T' is estimated to be 130K, well above 27K and well below 300K.}
\label{Fig3}
\end{figure}

In figure \ref{Fig1} we present a typical spectrum obtained at 5K.
One can distinguish a low energy feature (see lower red arrows). The
shape and the temperature dependence of this features allow us to
identify it with the superconducting gap. The shape of the feature
can be isolated in figure \ref{Fig2} (a) when dividing out the `U'
shaped background (red fit figure \ref{Fig1} named `low energy
fit'). As discussed further, this background choice is the best
representation of the spectra above $T_c$. By isolating the low
energy feature in figure \ref{Fig2} (a), one can then notice weak
coherence peaks in the spectrum. These peaks are incompatible with
the low energy pseudogap measured by tunnel junction measurements
\cite{Alff, DaganPG} but compatible with the signature of the
superconducting gap. The zero bias conductance (ZBC) value of 80\%
is slightly higher than other well studied cuprates such as YBCO
where the ZBC is generally found to be around 50\%. Using this
spectrum one finds the gap value from peak to peak to be
$2\Delta_{pp}$=18meV. In a BCS picture this corresponds to a gap of
$2\Delta$=12meV. Using this extraction method, tunneling spectra
measurements performed at various distant points on a same sample
and on two different optimally doped samples show a superconducting
gap $2\Delta$ ranging from 7meV to 12meV. These variations could be
explained by inhomogeneities on large scales in the oxygen content
arising in the reduction process and/or variation of cerium
concentration on the micron scale. Using these gap values we find a
ratio $2\Delta$/$k_BT_c$ varying from 4 to 7. Other spectroscopic
techniques find this ratio to be 4.4 from Raman \cite{Blumberg} in
NCCO, 2.3 from photoemission in PLCCO \cite{MatsuiSupra}, 5 from
optical conductivity of optimally doped PCCO \cite{ZimmersSupra} and
3.4 in NCCO and PLCCO from point contact tunneling spectroscopy
\cite{Shan}. All of these ratios are on the lower limit of the one
we find in SCCO. Figure \ref{Fig3} presents the temperature
dependence from 5K to 300K on a second optimally doped sample. When
raising the temperature through the superconducting transition from
5K to 27K, one observes that the low energy feature vanishes. All of
the arguments described above allow us to identify this low energy
feature with the superconducting gap.

We now discuss the high energy part of the spectra. Similar to the
analysis used to isolate the superconducting gap, we have adjusted
the conductance background above and below the high energy feature.
This background fit presented in blue in figure \ref{Fig1} is
divided out to give figure \ref{Fig2} (b). We have chosen the
simplest background line in order to isolate the characteristic
energy scale of the high energy feature. Note that such a 'U' shaped
background is also observed in many other superconducting cuprates
such as YBCO. The magnitude of the high energy feature is identified
to be $2\Delta$'=120meV (see blue arrows presented in figure
\ref{Fig1} and figure \ref{Fig2} (b)). This value is comparable in
magnitude with the large energy pseudogap measured by ARPES
\cite{ArmitageNCCO, ArmitageSCCO} and optical spectroscopy in NCCO
\cite{Onose}, PCCO \cite{Zimmers} and SCCO \cite{ArmitageSCCO}.
However, further quantitative comparison is hard to pursue since the
estimate of the large energy pseudogap maggnitude clearly depends on
the spectroscopy technique used. Indeed, a recent study on a unique
underdoped SCCO sample \cite{ArmitageSCCO} has shown that the
estimate of the large energy pseudogap varies from 200meV to 300meV
when measuring by photoemission or optical conductivity. The most
interesting and controversial aspect of this spectrum is the
observation of the superconducting gap and large energy pseudogap at
the same local measurement points. The resolution of the tip was
evaluated to be 10\,\AA. The vast majority of data recorded using
various samples showed both gaps in each spectrum. If the high
energy feature is identified to be the signature of an
antiferromagnetic order in the sample, this would then imply that
both superconductivity and antiferromagnetism could coexist locally.

\begin{figure}
\begin{center}
\includegraphics[width=9cm]{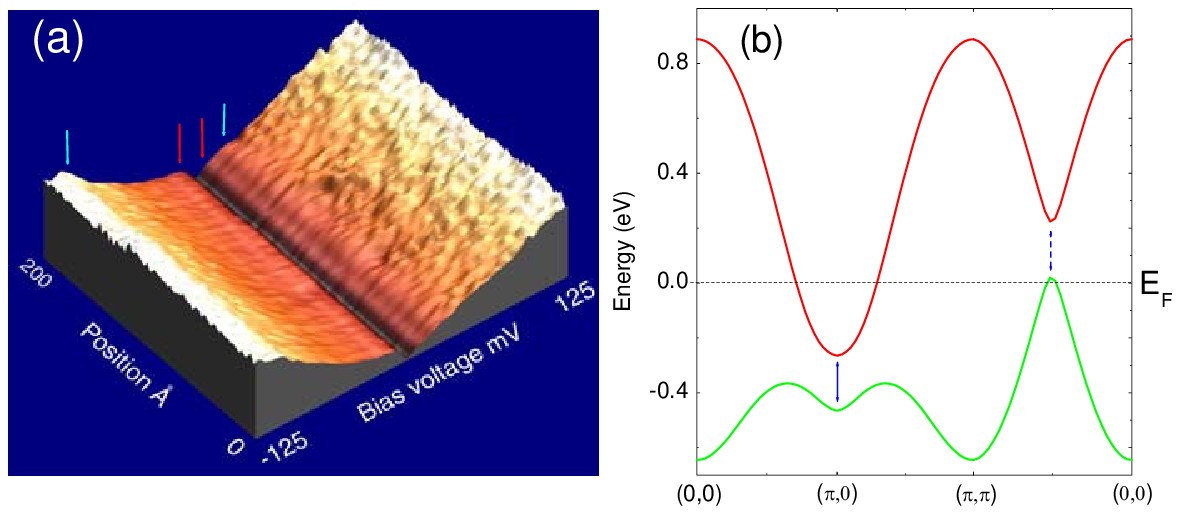}
\end{center}\caption{(Color online) (a) Tunneling spectra following a 200\AA\, line of SCCO
x=0.145 at 5K (plotted using WSxM \cite{WSxM}). The superconducting
gap is identified in the center. The high energy feature can clearly
be observed on the negative bias side  but only suggested at
positive bias (see blue arrows). As explained in figure (b) this
asymmetry relative to zero bias could be explained by the fact that
the tunnel current here is mostly sensitive to ($\pi$,0) or
(0,$\pi$) regions in these measurements. Finally one can notice that
both high and low energy features do not change with position.(b)
Band structure calculation of electron doped cuprate with a AF
order, using a spin density wave model with a 0.2eV gap
\cite{Zimmers, Millis}. This figure clearly illustrates how the
appearance of the AF gap in the tunnel spectra is influenced by the
weighted averaged around the Fermi surface. If the tunnel current is
mostly sensitive to the ($\pi$,0) or (0,$\pi$) regions, the AF gap
is shifted to negative bias (blue arrow). However, if the tunnel
current is mostly sensitive to the diagonal part of the Fermi
surface, the AF gap is shifted to positive bias (dashed blue arrow).
} \label{Fig4}
\end{figure}

The spectra shown in figure \ref{Fig1}(a) presents the advantage of
being symmetric with respect to the zero-bias voltage. This symmetry
is not always present in the data. Indeed, in many spectra the high
energy features were shifted to one or the other sides of the
zero-bias voltage (see figure \ref{Fig1} (b) and (c)). This
asymmetry could be understood using the following arguments:\\- The
measured tunnelling spectra result from averaging over the entire
Fermi surface. Contrary to tunneling in simple metal, in complex
materials such as SCCO, this averaging could be weighted in
preferential directions of $k$-space due to particular geometry
between the tip and the surface \cite{Footnote1}. If this tunneling
configuration changes while probing spatially distant points of a
sample, this should be followed by a variation in the weighting
factors.\\- In the particular case of SCCO, these variations could
generate important modifications of the tunneling spectra. Indeed,
when ordered antiferromagnetically, SCCO band structure is found to
be a two-band system near the Fermi energy \cite{ArmitageNCCO,
ArmitageSCCO} (see Figure \ref{Fig4} (b)). By definition, the gap
between both bands will appear to be centered with respect to the
Fermi energy if the tunneling current is most sensitive to the
nesting points (hot-spots). However, this gap will appear to be
off-centered and larger if the current is most sensitive to the
antinodal regions ($\pi$,0) or (0,$\pi$) (see blue arrow figure
\ref{Fig4}b) or the nodal region ($\pi$/2,$\pi$/2) direction (see
dashed blue arrow, figure \ref{Fig4}b).

We now turn to the temperature dependence of this high energy
feature. Figure \ref{Fig3} shows the SCCO spectra of a different
optimally doped sample x=0.155 (having a $T_c$ slightly lower than
the sample discussed previously in figure \ref{Fig1} and \ref{Fig2})
at three different temperatures, 5K, 27K and 300K. Considering
temperature drifts, the three spectra were taken in the same local
region and represent typical results obtained when raising the
temperature. At low temperatures one can again identify two energy
scales represented by the red and blue arrows. When passing above
the superconducting transition temperature, the superconducting gap
closes but the high energy feature remains. However, when raising
the temperature to room temperature, we observe the closing of this
energy gap, only leaving the overall `U' shaped line, similar to the
background fit used previously. Since the closing of the large
energy pseudogap has been shown to be smooth \cite{Zimmers}, it is
difficult to identify the characteristic temperature T' above which
this feature disappears in the conductance spectra. However, optical
measurements of NCCO and PCCO were able to establish the relation
$2\Delta'$/$k_BT'$=16 between the pseudogap energy $\Delta$' and the
gap characteristic opening temperature T'. Using this relation and
the value $2\Delta$'=90meV from figure \ref{Fig3}, one finds T'=130K
for this sample. This simple estimate of T' explains why one
observes the high energy gap at 5K and 27K but not at 300K.

Figure \ref{Fig4} (a) presents the tunnel spectra taken along a
200\AA\, line chosen from a topographic/spectrographic measurement
of a 330\AA$^2$ square. The superconducting gap is observed around
zero bias voltage (see red arrows). The high energy feature is
clearly seen on the negative bias voltage side and is only suggested
on the positive bias voltage side (see blue arrows)
\cite{Footnote2}. In this scenario, the high energy feature is found
to range from -100meV to +30meV making it comparable in energy (but
shifted towards negative bias) with respect to the high energy
feature presented in figure \ref{Fig1}(a) (both spectra were
measured in regions close by). Moreover this image shows that both
the high and low energy features are very stable in amplitude
throughout nanometer distances. The homogeneity of the
superconducting gap magnitude following a 35nm line scan was
previously reported in NCCO \cite{Kashiwaya}. This suggests that the
DOS map fluctuation observed in hole-doped cuprates \cite{Cren,
Davis} is not observed in electron-doped cuprates on the same scale.
The variation in the superconducting gap mentioned previously at
various distant points on the surface can be explained by slight
inhomogeneities in the oxygen content arising in the reduction
process.

We have reported the normal state and superconducting state local
tunneling spectra of SCCO. On a large scale, the superconducting
gap $\Delta$ is observed to range from 3.5meV to 6meV, the ratio
$2\Delta'$/$k_BT_c$ thus varying from 4 to 7. At low temperatures
a large energy feature in the tunneling spectra is observed. This
one remains above $T_c$ but disappears at higher temperatures. On
the nanometer scale the amplitudes of the superconducting gap and
the high energy feature are found not to vary with position. The
high energy feature could be the signature of the large energy
pseudogap observed by photoemission and optical spectroscopy
(interpreted as the signature of AF order). In this case these
results would suggest the coexistence on a local scale of
antiferromagnetism and superconductivity. Further measurements
need to be performed to confirm this interpretation.

This work was supported by NSF grants DMR-0352735, DMR-0303112 and
DMR-0645461. A.Z. acknowledges Fellowship support from the ICAM
International Materials Institute.


\begin{thebibliography}{00}
\bibitem{Footnote1}
A study of all possible reasons leading to such asymmetries will
be pursued in the future.
\bibitem{Footnote2}
In some particular line scans, a probable Coulomb blockage is
occasionally observed at higher energy but changes with position
making it unrelated to the other spectroscopic features mentioned.
\end{thebibliography}
\end{document}